\documentclass[a4paper,12pt]{article}

\begin{document}
\topmargin 0pt \oddsidemargin 0mm

\renewcommand{\thefootnote}{\fnsymbol{footnote}}
\begin{titlepage}
\begin{flushright}
 hep-th/0311240
\end{flushright}

\vspace{5mm}
\begin{center}
{\Large \bf A Note on Thermodynamics of Black Holes in Lovelock
Gravity} \vspace{32mm}

{\large Rong-Gen Cai\footnote{e-mail address: cairg@itp.ac.cn}}

\vspace{10mm} {\em  Institute of Theoretical Physics, Chinese
Academy of Sciences,\\
 P.O. Box 2735, Beijing 100080, China }

\end{center}

\vspace{5mm} \centerline{{\bf{Abstract}}}
 \vspace{5mm}
The Lovelock gravity consists of the dimensionally extended Euler
densities. The geometry and horizon structure of black hole
solutions could be quite complicated in this gravity, however, we
find that some thermodynamic quantities of the black holes like
the mass, Hawking temperature and entropy, have simple forms
expressed in terms of horizon radius. The case with black hole
horizon being a Ricci flat hypersurface is particularly  simple.
In that case the black holes  are always thermodynamically stable
with a positive heat capacity and their entropy still obeys the
area formula, which is no longer valid for black holes with
positive or negative constant curvature horizon hypersurface. In
addition, for black holes in the gravity theory of Ricci scalar
plus a $2n$-dimensional Euler density with a positive coefficient,
thermodynamically stable small black holes always exist in
$D=2n+1$ dimensions, which are absent in the case without the
Euler density term, while the thermodynamic properties of the
black hole solutions with the Euler density term are qualitatively
similar to those of black holes without the Euler density term as
$D>2n+1$.

\end{titlepage}

\newpage
\renewcommand{\thefootnote}{\arabic{footnote}}
\setcounter{footnote}{0} \setcounter{page}{2}


 Over the past years  gravity with higher derivative curvature
 terms has received a lot of attention, in particular, in the
 brane world scenario and black hole thermodynamics. In the former,
 the motivation is to study the corrections of these higher derivative terms
 to the Newton law of gravity on the brane, or to avoid the singularity in bulk
 by using these higher derivative terms. In the field of black
 hole thermodynamics,  it is expected to gain some insights into
 quantum gravity since the thermodynamic property of black hole is
 essentially a quantum feature of gravity. On the other hand, it is
 possible through the thermodynamics of black holes with AdS asymptotic
  to study the thermodynamic
 properties and phase
 structure of a ceratin field theory because due to the AdS/CFT
 correspondence, some higher derivative curvature terms can be
 regarded as the corrections of large $N$ expansion of dual field
 theory.

 Among the higher derivative gravity theories, the so-called Lovelock
 gravity~\cite{Love} is quite special, whose
 Lagrangian  consists of the dimensionally
 extended Euler densities
 \begin{equation}
 \label{eq1}
 {\cal L}= \sum^m_{n=0}c_n{\cal L}_n,
\end{equation}
where $c_n$ is an arbitrary constant and ${\cal L}_n $ is the
Euler density of a $2n$-dimensional manifold:
\begin{equation}
\label{eq2}
 {\cal L}_n= 2^{-n} \delta^{a_1b_1\cdots a_nb_n}_{c_1d_1\cdots
 c_nd_n}R^{c_1d_1}_{~~~a_1b_1}\cdots R^{c_nd_n}_{~~~a_nb_n}.
 \end{equation}
 Here the generalized delta function is totally antisymmetric in
 both sets of indices. ${\cal L}_0$ is set to one, the constant
 $c_0$ is therefore proportional to the cosmological constant. ${\cal L}_1$ gives
 us the usual curvature scalar term. In order the Einstein's general
 relativity to be recovered in the low energy limit, the constant $c_1$ must
 be positive. Here, for simplicity, we just set the constant
 $c_1=1$. The ${\cal L}_2$ term is the Gauss-Bonnet one, which
 often appears in the recent literature. Except for the
 advantage that the equations of motion of the Lovelock gravity,
 as the case of the Einstein's general relativity, do not contain
 terms with more than second derivatives of metric, the Lovelock
 gravity has been shown to be free of ghost when expanding on a
 flat space, evading any problems with unitarity~\cite{BouDes}.
 Here it should be stressed that although the Lagrangian
 (\ref{eq1}) consists of some higher derivative curvature terms,
 the Lovelock gravity is not essentially a higher derivative
 gravity since its equations of motion do not contain terms with
 more than second derivatives of metric. Just due to this, the
 Lovelock gravity is free of ghost~\cite{Zwiebach}.

 In the literature concerning on the Lovelock garvity, the
 extensively
  studied is the so-called Gauss-Bonnet gravity, whose Lagrangian is the sum of
  the curvature scalar term ${\cal L}_ 1$ and the Gauss-Bonnet term
  ${\cal L}_2$, the Euler density of a $4$-dimensional manifold.
  Sometimes, a cosmological constant is added to the Lagrangian.
  In this theory, the static, spherically symmetric black hole
  solution was found in Refs.~\cite{BouDes,Whee}. The black hole
  solutions
  with nontrivial horizon topology were studied in Ref.~\cite{Cai1}.
  Refs.~\cite{Cai2,GP} discussed some
  aspects of holography of the Gauss-Bonnet gravity.  In
  particular, it is worth mentioning here that with a positive
  Gauss-Bonnet coefficient $c_2$, in spite of the asymptotic
  behavior (asymptotically (anti-)de Sitter~\cite{Cai1,Cai3} or
  flat~\cite{MyerSim}) of black hole
  solutions, it is found that a locally stable small black hole always appears when
  the spacetime dimension $D=5$, which is absent in the case without the
  Gauss-Bonnet term, while $D \ge 6$, the thermodynamic behavior
  of the Gauss-Bonnet black hole is qualitatively similar to the case
  without the Gauss-Bonnet term (see also related discussions
  in \cite{Nojiri,Nep}).

 The Lagrangian (\ref{eq1}) looks complicated. It is therefore
 a little surprise to know that the static, spherically
 symmetric solution can be found in the sense that the metric
 function is determined by solving for a real root of a polynomial
 equation~\cite{Whee}. Since the gravity ({\ref{eq1}) includes many arbitrary
 coefficients $c_n$, it is not an easy matter to extract physical
 information from the solution. In Refs.~\cite{Ban,Cri} by choosing a special
 set of coefficients, the metric function can be expressed in a
 simple form. These solutions could be explained as spherically symmetric
 black hole solutions. Black hole solutions with nontrivial
 horizon topology in this gravity with those special coefficients
 have also been studied in Refs.~\cite{Cai4,Aro}.

For the general case with arbitrary coefficients $c_n$, the
static, spherically symmetric solution was found in
Refs.~\cite{Whee,MyerSim}
\begin{equation}
\label{eq3}
 ds^2= -f(r) dt^2 +f(r)^{-1}dr^2 + r^2 d\Omega_D^2,
 \end{equation}
 where $d\Omega_D^2$ denotes the line element of an
 $(D-2)$-dimensional unit sphere and the metric function $f(r)$
 is given by
 \begin{equation}
 \label{eq4}
 f(r) = 1 -r^2 F(r).
 \end{equation}
$F(r)$ is determined by solving for real roots of the following
$m$th-order polynomial equation
\begin{equation}
\label{eq5}
 \sum^m_{n=0}\hat c_n F^n(r)=\frac{16\pi G
M}{(D-2)\Omega_D r^{D-1}}.
\end{equation}
Here $G$ is the Newton constant in $D$ dimensions,
$\Omega_D=2\pi^{(D-2)/2}/\Gamma[(D-2)/2]$ is  the volume of an
$(D-2)$-dimensional unit sphere, $M$ is an integration constant,
and the coefficient $\hat c_n$ is given by
\begin{eqnarray}
&& \hat c_0= \frac{c_0}{(D-1)(D-2)}, ~~~~~~~ \hat c_1=1,
\nonumber\\
&& \hat c_n =c_n \Pi ^{2m}_{i=3}(D-i) ~~~~~~{\rm for\ } n>1.
\end{eqnarray}
The asymptotic behavior and causal structure of the solution have
been analyzed in detail by Myers and Simon in Ref.~\cite{MyerSim}.
It is found that even when the cosmological constant $c_0$
vanishes, the solution could be asymptotically de Sitter, flat, or
anti-de Sitter, which depends on the coefficients $c_n$,(in other
words, it can be seen from (\ref{eq4}) that  the solution is
asymptotically de Sitter, flat, or anti-de Sitter if $F(r)$
approaches to a positive constant, zero or a negative constant as
$r\to \infty$, respectively) and that (black hole/cosmological)
horizon structure is quite rich. But we do not repeat them here.
Further it is easy to conclude that the integration constant $M$
is the ADM mass when the solution is asymptotically flat, while it
corresponds to the AD mass for the asymptotically (anti-)de Sitter
case~\cite{AD}.

Nowadays it is well-known that in asymptotically anti-de Sitter
space, the black hole horizon could be topologically nontrivial:
the horizon  can be a closed hypersurface with positive, zero, or
negative constant curvature~\cite{Topo}. Such black holes are
called topological black holes. Now we generalize the spherically
symmetric solution (\ref{eq3}) to more general case:
\begin{equation}
\label{eq7}
 ds^2=-f (r) dt^2 + f(r)^{-1}dr^2 + r^2 d\Sigma_D^2,
 \end{equation}
 where $d\Sigma_D$ denotes a line element of an $(D-2)$-dimensional
 hypersurface with constant scalar curvature $(D-2)(D-3)k$ and
 volume $\Sigma_D$. Here $k$ is a constant. Without loss of
 generality, $k$ can be set to $\pm 1$ or zero. In this case, the metric
 function $f(r)$ becomes
 \begin{equation}
 \label{eq8}
 f(r) =k -r^2 F(r),
 \end{equation}
 and $F(r)$ still obeys the equation (\ref{eq5}) with $\Omega_D$
 replaced by $\Sigma_D$. Following Myers and Simon~\cite{MyerSim},
  we can also make
 detailed analysis for the solution ({\ref{eq8}) on its the horizon
 structure. Note that only when the solution is asymptotically
 anti-de Sitter, black hole horizon will appear for any $k$; when
 the solution is asymptotical flat, it is possible to have black
 hole horizon only for the case $k=1$; when the solution is
 asymptotically de Sitter, a cosmological horizon appears, of course,
 the black hole may also occur for $k=1$ in this case.
 When the solution is asymptotically de Sitter, the
 solution (\ref{eq8}) with any $k$ is the generalization of the topological de
 Sitter spaces introduced in \cite{CMZ}.

 For our purpose, without any detailed analysis, we just assume
 that a black hole horizon exists for the solution (\ref{eq7}).
 Although the solution (\ref{eq7}) looks involved, we will show that
 some thermodynamic quantities associated with the black hole
 can have simple expressions in terms of horizon radius. According
 to the metric (\ref{eq7}), the black hole horizon radius $r_+$
 is determined via the equation $f(r_+)=0$. Due to the equation (\ref{eq8}),
 one has
 \begin{equation}
 \label{eq9}
 r_+^2 =k/F(r_+).
 \end{equation}
 Note that when $k=0$, it corresponds to $r_+=0$ or $F(r_+)=0$.
 For the former case, it indicates that the black hole horizon
 coincides with the singularity at $r=0$. The Hawking temperature
 of the black hole can be obtained by using the periodicity of
 imaginary time in the metric. Continuing the black hole solution to
 its  Euclidean section via $\tau =i t$, the resulting  manifold will
 have a conical singularity at the black hole horizon $r_+$ if the period $\beta$
 of the Euclidean time $\tau $ is arbitrary. To remove the conical
 singularity, the period must be fixed to a special value.  The
 periodicity of the Euclidean time appears in the quantum field's Euclidean propagator
 when one considers a certain quantum field in the black hole
 background. In quantum field theory at finite temperature,
 the period of the Euclidean time is explained as the inverse
 temperature, which is just the inverse Hawking temperature of black hole.
 For the black hole solution (\ref{eq7}), the special value of
 the period of the Euclidean time is found to be
 \begin{equation}
 \label{eq10}
 \beta \equiv 1/T =4\pi/ f'(r)|_{r=r_+},
 \end{equation}
 where a prime denotes the derivative with respect to $r$. That
 is, the Hawking temperature is
 \begin{equation}
 \label{eq11}
 T = \frac{1}{4\pi}f'(r)|_{r=r_+}= -\frac{1}{4\pi}(2k/
 r_+ +r^2_+ F'(r)|_{r=r_+}).
 \end{equation}
 Here we have used the relation (\ref{eq9}).
 To get a simplified expression, we have from the equation
 (\ref{eq5}) the mass of black hole in terms of the horizon
 radius
 \begin{eqnarray}
 \label{eq12}
 M &=&\frac{(D-2)\Sigma_D r_+^{D-1}}{16\pi G}\sum^m_{n=0}\hat c_n
  F^n(r_+) \nonumber \\
  &=&\frac{(D-2)\Sigma_D r_+^{D-1}}{16\pi G}\sum^m_{n=0}\hat c_n
  (kr_+^{-2})^n.
 \end{eqnarray}
 When $k=0$, it reduces to
 \begin{equation}
 \label{eq13}
 M_{k=0}=\frac{(D-2)\Sigma_D r_+^{D-1}}{16\pi G}\hat c_0.
 \end{equation}
 To obtain $F'(r)|_{r=r_+}$, taking derivatives on both sides of
 equation (\ref{eq5}) with respect to $r$ and using (\ref{eq12}) and
 (\ref{eq9}), we then get
 \begin{equation}
 \label{eq14}
 F'(r)|_{r=r_+}=-\frac{(D-1)\sum^m_{n=0}\hat c_n (kr_+^{-2})^n}
    {r_+\sum^m_{n=1}n\hat c_n (kr_+^{-2})^{n-1}}.
 \end{equation}
 Substituting into (\ref{eq11}), we reach the expression of
 the Hawking temperature
 \begin{equation}
 \label{eq15}
 T=\frac{\sum^m_{n=0}(D-2n-1)\hat c_n k (kr_+ ^{-2})^{n-1}}
    {4\pi r_+ \sum^m_{n=1}n \hat c_n (kr_+^{-2})^{n-1}}.
    \end{equation}
  When $k=0$, we find a very simple expression
  \begin{equation}
  \label{eq16}
  T_{k=0}=\frac{(D-1) \hat c_0}{4\pi}r_+,
  \end{equation}
which is remarkable result: the Hawking temperature does not
explicitly depend on other constants $\hat c_n (n>1)$.

 Another important thermodynamic quantity associated with black hole
horizon is its entropy. Black hole behaves as a thermodynamic
system, its thermodynamic quantities must obey the first law of
thermodynamics, $dM =TdS$.  Using this relation, in
Ref.~\cite{Cai4} we have derived the black hole entropy in a
higher derivative gravity theory, and in \cite{Cai1} we have
obtained the same entropy of Gauss-Bonnet black holes as the
Euclidean approach~\cite{MyerSim}. Here we use the first law to
get the entropy of black hole (\ref{eq7}). Integrating the first
law, we have
\begin{equation}
\label{eq17}
 S=\int T^{-1}dM = \int^{r_+}_0 T^{-1}\frac{\partial M}{\partial
 r_+}dr_+.
 \end{equation}
 Here we have assumed that the entropy vanishes when the horizon
 radius shrinks to zero. Thus once the Hawking temperature and the
 mass are given in terms of the horizon radius, one can obtain the
 entropy of black hole using (\ref{eq17}). Substituting
 (\ref{eq12}) and (\ref{eq16}) into (\ref{eq17}), we arrive at
 \begin{equation}
 \label{eq18}
 S =\frac{\Sigma_D r_+^{D-2}}{4G}\sum^m_{n=1}\frac{n(D-2)}{D-2n}
    \hat c_n (kr_+^{-2})^{n-1}.
    \end{equation}
Once again, the case with $k=0$ is very special. In that case, one
can see from (\ref{eq18}) that there is only one term with $n=1$
has the contribution to the entropy:
\begin{equation}
\label{eq19}
S_{k=0}=\frac{\Sigma_D r_+^{D-2}}{4G}.
\end{equation}
Note that $\hat c_1=1$ and $\Sigma_D r_+^{D-2}$ is just the
horizon area of black hole. We therefore conclude that in spite of
the higher derivative terms, the entropy of black holes with $k=0$
always obeys the area formula of black hole entropy. For other
cases with $k=\pm 1$, the area formula of black hole entropy  does
no longer hold obviously.

Some remarks are in order here. First we notice that although the
asymptotic behavior and horizon structure of the black hole
solution (\ref{eq7}) are complex, their thermodynamic quantities
have simple expressions in terms of horizon radius. Their mass,
Hawking temperature and entropy are given by (\ref{eq12}),
(\ref{eq15}) and (\ref{eq18}), respectively. Second, when the
horizon is a Ricci flat hypersurface, namely $k=0$, the
thermodynamic quantities of the black hole have quite simple forms
given by (\ref{eq13}), (\ref{eq16}) and (\ref{eq19}),
respectively. Further, from the relations (\ref{eq13}) and
(\ref{eq16}) of mass and Hawking temperature to the cosmological
constant $\hat c_0$, one has to have $\hat c_0>0$, a negative
cosmological constant, in order to make these relations sense. In
addition we see  from (\ref{eq13}) and (\ref{eq16}) that the black
holes with $k=0$ are always thermodynamically stable with positive
heat capacity.

Third, as mentioned above, in spite of the asymptotic behavior of
the Gauss-Bonnet black hole solution, the Gauss-Bonnet black holes
with positive Gauss-Bonnet coefficient $\hat c_2$ always have a
thermodynamically stable branch with small horizon radius in $D=5$
dimensions, while their thermodynamic properties are qualitatively
similar to the case without the Gauss-Bonnet term if $D\ge 6$.
Here we show that this feature persists for gravity with higher
dimensional Euler density. For example, let us consider a gravity
theory consisting of a cosmological constant term $\hat c_0$, a
curvature scalar term $R$ and a $2n$-dimensional Euler density
${\cal L}_n$.  From (\ref{eq15}), we have the Hawking temperature
of the black hole
\begin{equation}
\label{eq20}
 T=\frac{(D-1)\hat c_0 r_+^{2n} +(D-3) r_+^{2n-2} +(D-2n-1)\hat
 c_n}{4\pi r_+(r_+^{2n-2}+ n\hat c_n)},
 \end{equation}
 where we have already set $k=1$. Suppose $\hat c_n>0$, which
 makes the horizon radius $r_+$ have minimal value $r_+=0$
 (cf.~\cite{Cai1,Cai3,MyerSim}),
 we can easily see that the behavior of the Hawking temperature
 crucially  depends on spacetime dimension. When $D=2n+1$,  the
 Hawking temperature always increases monotonically from $T=0$ at
 $r_+=0$ for small horizon radius, independent of the cosmological
 constant $\hat c_0$.  This is consistent with the fact that at
 much smaller scale than the cosmological radius $1/\sqrt{|\hat c_0|}$,
 (if it does not vanishes), the cosmological constant has a negligible
 effect on physics on that scale. Therefore in this case the small black hole
 is thermodynamically stable with positive heat capacity. Of course,
 for larger black holes, the behavior of Hawking temperature
 depends on the asymptotic behavior of the black hole solutions.
 From (\ref{eq20}) it can be seen that the effect of the
 coefficient $\hat c_n$ is small when $r_+^{2n-2} > n \hat
 c_n$. On the other hand, when $D>2n+1$, the Hawking temperature
 always decreases monotonically from infinity at $r_+=0$ for small
 black holes, which implies that the heat capacity is negative, as
 the case without the Euler density term. For larger horizon radius,
 the effect of the Euler density term is once again small.  Therefore
 the thermodynamic behavior of these black holes is qualitatively
 same as the case without the Euler density term as $D>2n+1$.

Finally we see from the entropy (\ref{eq18}) that the first term
is just quite familiar area term of black hole horizon, other
terms comes from contributions of higher dimensional Euler
densities. In this expression the cosmological constant term does
not appear explicitly. This is an expected result since entropy of
black hole is a function of horizon geometry~\cite{Wald}. Here we
see that horizon topology also plays an important role for entropy
of black holes in gravity with higher derivative terms. To see
further the feature that black hole entropy is a character of
horizon geometry and topology, let us add a Maxwell field to the
Lovelock gravity (\ref{eq1}).  We will see that entropy of the
charged black hole in Lovelock gravity still have the expression
(\ref{eq18}). That is, the electric charge $q$ does not appear
explicitly in the black hole entropy expressed in terms of horizon
radius. When a Maxwell field is present, we have a charged black
hole solution with metric (\ref{eq7}). Here metric function $f(r)$
is still given by (\ref{eq8}), but $F(r)$ has to
satisfy~\cite{MyerSim,Wil}
\begin{equation}
\label{eq21} \sum^m_{n=0}\hat c_n F^n(r) =\frac{16\pi G M}{(D-2)
\Sigma_D r^{D-1}} - \frac{q^2}{r^{2D-4}}.
\end{equation}
In this case, black hole horizon $r_+$ is still determined by the
equation $f(r_+)=0$. So the mass of black hole can be expressed in
terms of horizon radius $r_+$ and charge $q$
\begin{equation}
\label{eq22} M=\frac{(D-2)\Sigma_D r^{D-1}_+}{16\pi G} \left (
 \sum^m_{n=0}\hat c_n (k r_+^{-2})^n +
 \frac{q^2}{r_+^{2D-4}}\right).
 \end{equation}
 The Hawking temperature is found to be
 \begin{equation}
 \label{eq23}
 T=\frac{\sum^m_{n=0} (D-2n-1) k \hat c_n (k r_+^{-2})^{n-1}-
   (D-3) q^2/r_+^{2D-6}}{4\pi r_+ \sum^m_{n=1} n\hat c_n
   (kr_+^{-2})^{n-1}}.
\end{equation}
The variation of the mass (\ref{eq22}) with respect to the horizon
radius $r_+$ is
\begin{equation}
\label{eq24} \left (\frac{\partial M}{\partial r_+}\right)_q
 =\frac{(D-2)\Sigma_D r^{D-4}_+}{16\pi G}\left (\sum^m_{n=0}
  (D-2n-1) k \hat c_n (k r_+^{-2})^{n-1}-
   (D-3) q^2/r_+^{2D-6}\right).
 \end{equation}
 Using (\ref{eq17}) once again, and keeping $q$ as a constant in
 calculation, we get
 \begin{equation}
 \label{eq25}
S =\frac{\Sigma_D r_+^{D-2}}{4G}\sum^m_{n=1}\frac{n(D-2)}{D-2n}
    \hat c_n (kr_+^{-2})^{n-1}.
    \end{equation}
It has a same form as the entropy (\ref{eq18}) of a uncharged
black hole.

In summary we have first generalized the static, spherically
symmetric black hole solution in Lovelock gravity to the case
where black hole horizon can be a positive, zero or negative
constant curvature hypersurface. Although the geometry and horizon
structure of the black hole solution could be quite complicated,
in terms of horizon radius, we have found that some thermodynamic
quantities like the black hole mass, Hawking temperature and
entropy, have simple expressions. In particular, the case with
Ricci flat horizon is remarkably simple: these black holes are
thermodynamically stable with a positive heat capacity and their
entropy always obeys the horizon area formula. By explicit
calculation, it has been shown that black hole entropy depends on
not only the horizon geometry, but also the horizon topology
structure. In addition, the feature has been found to be universal
that for black hole solutions in gravity of Ricci scalar plus a
$2n$-dimensional dimensional Euler density, when $D=2n+1$, the
thermodynamically stable small black holes always appear with a
positive heat capacity, which are absent in the case without the
Euler density term. In $D>2n+1$, however, the thermodynamic
properties of black holes become qualitatively similar to those of
black holes without the Euler density term. This generalized the
discussions of Gauss-Bonnet black holes to a more general case.

\section*{Acknowledgments}
This work was supported in part by a grant from Chinese Academy of
Sciences, a grant from from NSFC, a grant from the Ministry of
Education of China, and by the Ministry of Science and Technology
of China under grant No. TG1999075401.

\end{document}